\documentclass{svproc}

\usepackage{graphicx}
\usepackage[table]{xcolor}
\usepackage{url}

\arrayrulecolor[HTML]{DB5800}

\begin{document}
\mainmatter              

\title{A Lightweight Encryption Scheme for IoT Devices in the Fog\\}

\titlerunning{Lightweight IoT Encryption}  
\author{Matthew Chun\inst{1} \and Stefan Weber\inst{2} \and Hitesh Tewari\inst{3}}
\authorrunning{Matthew Chun et al.} 
%
\tocauthor{Matthew Chun, Stefan Weber, Hitesh Tewari}
\institute{Amherst College, MA, USA\\
\email{machun24@amherst.edu}\\ 
Trinity College Dublin, Ireland\\
\and \email{sweber@tcd.ie}\\ 
\and \email{htewari@tcd.ie}
}

\maketitle              

\begin{abstract}
The Internet of Things (IoT) is the collection of everyday smart devices which connect to the Cloud, often through Fog nodes, to transmit and receive information. These everyday devices are distinct from traditional computers because they typically have notable constraints on their RAM, flash memory, and computational power. Due to these constraints, we believe that many of the proposed encryption schemes are too heavyweight to be employed in the IoT. In this paper we present a lightweight, flexible encryption scheme that relies on the one-way information loss property of a secure hash function. Our scheme imposes minimal computational and storage requirements, and imposes no non-negligible burdens on the encrypting device, except for the hash itself. We find that the encryption algorithm is particularly lightweight, and holds up strongly in terms of its speed and memory efficiency.
\end{abstract}

\section{Introduction}

In this paper, we present an encryption protocol intended specifically for deployment into IoT devices with constrained resources. Possible applications for this encryption scheme include traffic lights, parking meters, homemade electronics, and other small smart appliances. Our encryption scheme uses a secure hash function to create keys. The primary benefit for choosing to base our scheme on a secure hash function is that it causes the hash function, often with well-known quantities such as encryption/decryption efficiency and RAM requirements, to dominate the encryption and decryption process. Furthermore, as new secure hash functions continue to be created, the framework presented with this simple encryption scheme can see an improvement in performance simply by replacing our provided example hash function with newer and more efficient ones. Hash functions can also be chosen for more or less security to balance against their required computational resources – meaning that an encryption framework using a hash function may be a lucrative option for a manufacturer that wants to fine-tune its encryption needs to precisely balance against the availability of computational resources. Though encryption protocols always provide confidentiality to a user’s data, our encryption protocol also provides in-built data integrity and authenticity.

The remainder of this paper is organized as follows. First, we examine existing encryption schemes that have been proposed for IoT environments and motivating innovations for our protocol. We then provide an overview of our system, detailing the network configuration within which our IoT devices are located, the key generation process, the data packet format we employ, and finally the encryption mechanism. We substantiate that the encryption protocol maintains confidentiality, integrity, and authenticity against standard cryptographic attacks. We also substantiate that the encryption framework is as minimal and lightweight as possible. We then provide the reader with experimental data displaying the efficiency of our encryption protocol, measured using a variety of Arduino-compatible devices. We analyze the experimental data and close with a few concluding remarks.

\section{Related Work}
There are numerous papers in the literature that survey the security risks to the IoT \cite{Neshenko},\cite{Alaba},\cite{Mosenia}. Risks such as battery draining, hardware trojans, malicious software updates, eavesdropping, denial-of-service, routing attacks etc. are just some of the areas that have been identified. These surveys also highlight the various approaches that researchers have employed in relation to developing lightweight cryptographic solutions for location privacy, authentication and encryption of data \cite {Buchanan},\cite{Latif}. Many of the proposals make use of symmetric key encryption algorithms such as AES \cite{EPS}. Others make use of asymmetric key algorithms such as RSA for encryption, authentication and non-repudiation. While some make use of a combination of symmetric and asymmetric key algorithms \cite{Adeel}. A number of schemes have also been proposed based on attribute-based encryption (ABE) techniques \cite{Yao}. Others make use of more cutting edge algorithms such as homomorphic encryption. While some advocate the use of secure hardware such as smart cards. Some schemes even propose offloading highly consuming cryptographic primitives to third parties \cite{Abdmeziem}.

We believe that the above approaches are still too heavyweight in terms of computational and communication costs, for them to be of any practical use in IoT and Fog environments. In \cite{Balasch}, the authors benchmark various hash algorithms on the ATtiny45 8-bit RISC microcontroller. The study shows that the lightweight hash algorithms outperform traditional hashing algorithms in terms of RAM memory and code size. Yet, to the best of our knowledge, a scheme similar to the one based on hash functions that we present in this paper has not been previously proposed for deployment in the IoT.

In \cite{SIMILAR}, the authors propose a very minimal encryption scheme using hashes that bears similarities to our current work. However, we believe the authors’ proposed variations on their scheme do not substantially increase its security. In contrast, we have added in-built authenticity and integrity while retaining the minimalist and efficient nature of their scheme. In \cite{Hashwork} the authors propose a hash algorithm specifically designed for the IoT, providing a collision-free secure hash option for encryption protocols. It is because of innovations such as these that we wish to propose an encryption protocol whose efficiency and security is controlled by the hash it uses. It is also because of innovations such as these that we strongly believe in the future of cryptographic hashing as a widespread tool for IoT encryption.

\section{System Overview}
We draw upon our knowledge of hash functions, computer networks, and their packet formats, in order to design a lightweight encryption scheme for the IoT. Our scheme makes use of a single, secure hashing algorithm, from which we can derive the key material. We make use of the fast XOR function to generate the ciphertext, and employ a well-known packet format technique to ensure the integrity of the data packets exchanged between the various entities in the system. Our protocol sidesteps the requirement for a secure channel to be established in order to exchange the key material.

\begin{figure}
\centerline{\includegraphics[width=20pc]{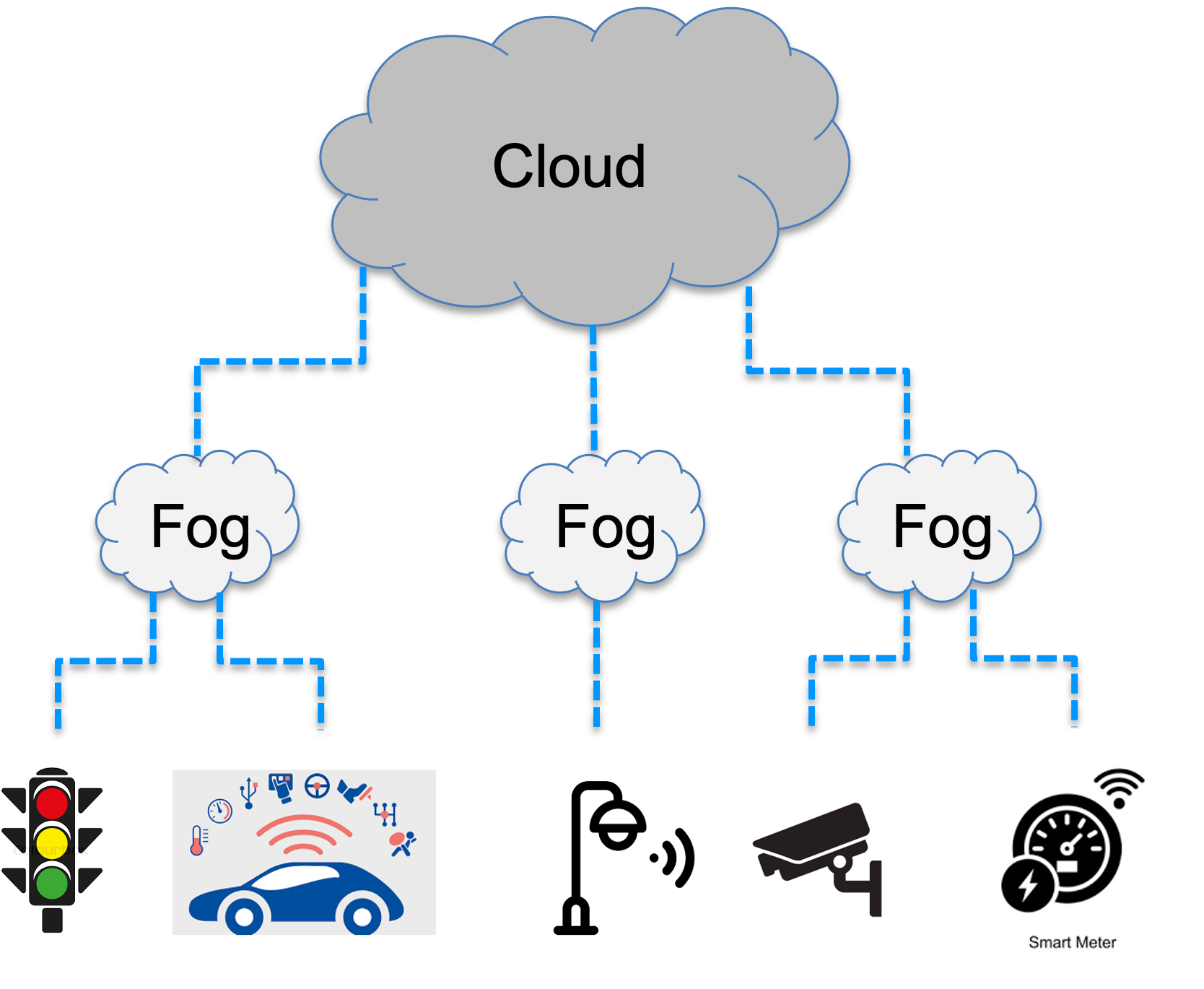}}
\caption{Coexistence of IoT, Fog and Cloud Environments}\label{fig:Fog}
\end{figure}

Figure \ref{fig:Fog} depicts a typical network configuration for our proposed system, where IoT devices are connected to the Cloud via Fog nodes. Fog computing \cite{Fog} is a new paradigm that extends the Cloud to the network edge. Fog nodes allow for storage, processing and retrieval of data to take place close to where it may be required, as opposed to storing it deep inside the network core. For example, it is faster and more efficient to store and process the telemetry data from vehicles in Fog nodes located close to the highway where they are travelling, as opposed to sending the data all the way to a Cloud server, and the subsequent retrieval of the data from the Cloud by another vehicle after a relatively short interval of time.

\subsection{Lightweight Encryption Protocol}
For our protocol, we are using an arbitrary 32-byte hash function ($H$). However, any hash function can be used in this framework, so long as it maps $\{0,1\}^{32} \mapsto \{0,1\}^{32}$ or $\{0,1\}^{64} \mapsto \{0,1\}^{64}$. Because our messages are fixed length by default, we do not require the hashes to map from $\{0,1\}^{n}$ for any $n$ less than the dimension of the codomain. Additionally, the hash function should have confusion and diffusion properties that are considered sufficiently secure for the application the encryption protocol is to be installed onto. 

For this breakdown, consider two communicating parties, such as an IoT device and a Fog node. Suppose one of the parties is about to encrypt and send a message, while the other is about to receive that message and decrypt it. We refer to the sender as $S$ and the receiver as $R$. Our protocol creates unique keys for each encryption by hashing a copy of a pre-deployed 27-byte shared secret key ($SK$) concatenated with a small 5-byte counter ($CTR$) that increments upon each use. The hash operation produces a 32-byte key that can be concatenated with itself to create a 64-byte key, which $S$ XORs with the message ($M$) to produce the encrypted message ($Enc_M$).

It is imperative that both $SK$ and $CTR$ be stored in non-volatile memory, as they are created before deployment, not at encryption time. 

\subsubsection{Key Setup:} \label{Key Setup}
We first embed the 27-byte shared secret $SK$ into both parties prior to deployment of the IoT device into the Fog. Then, $S$ creates a 27-byte copy of $SK$ known as $SK'$, stored as a local variable, which will be the object $S$ hashes during encryption, so $SK$ itself is never lost. We then concatenate a 5-byte (40-bit) counter $CTR$ to $SK'$ that starts at $000\ldots 001$ and increments each time data is encrypted or decrypted. We denote this concatenated product as $(SK'\parallel CTR)$. 

\subsubsection{Message Setup:}
The message $M$ sent from $S$ to $R$ must be sent in the following 64-byte format:

\begin{equation}
M = 55~bytes~data \parallel 1~byte~length \parallel 8~bytes~integrity~check
\end{equation}

\noindent This integrity check value must be equal to the first 8-bytes of $SK$, and will be checked to ensure integrity and authenticity upon decryption. The length communicates to $R$ how many of the first 55-bytes contain data, and how many of those bytes are padding.

\subsubsection{Encryption:}
To initiate the encryption process, $S$ increments its counter $CTR_S$. It then inputs $(SK'\parallel CTR_S)$ into the hash function $H$ and concatenates the output with a copy of itself to produce a 64-byte quantity as depicted in Equation \ref{eq:2}:

\begin{equation} \label{eq:2}
H(SK'\parallel CTR_S)~\parallel~H(SK'\parallel CTR_S)
\end{equation}

\noindent The sender $S$ then computes a bitwise XOR between (\ref{eq:2}) and $M$ to produce the encrypted message ($Enc_M$), which will be transmitted to $R$.

\begin{equation} \label{eq:3}
Enc_M = M~\oplus~(H(SK'\parallel CTR_S)~\parallel~H(SK'\parallel CTR_S))
\label{eq3}
\end{equation}

\noindent Since servers and Fog nodes communicate with many devices in parallel, if $R$ is a server or Fog node, $S$ transmits the following \textit{Tuple} to the receiver instead of just $Enc_M$:

\begin{equation}
Tuple = (ID,~Enc_M)
\end{equation}

\noindent where $ID$ is a node identifier, a MAC Address, or any other unique identifier of the sender sent in plaintext. By transmitting $ID$, $R$ can determine which $SK$ and $CTR$ values to use, as only one of $R$'s counters will be synchronized with $S$.

\subsubsection{Decryption:}
The receiving device receives $Enc_M$ given by (\ref{eq:3}). It first increments its own counter $CTR_R$ and tries to computes the original message.

\begin{equation}
M~=~Enc_M~\oplus~(H(SK'\parallel CTR_R)~\parallel~H(SK'\parallel CTR_R))
\end{equation}

\noindent Upon decrypting $Enc_M$, the device checks if the 8-byte integrity check in $M$ is equal to the first 8 bytes of $SK$. If they are not equal, the message is discarded and $CTR_R$ is decremented. This check ensures detection of active attacks to the messages, while demonstrating authenticity of the sender via knowledge of the shared secret $SK$. Due to the ``avalanche effect" of a secure hash, changing even a single bit will cause the integrity check to fail. Because $SK$ is never deployed outside either device, an illegitimate party would only be able to send a correctly decrypted illegitimate message if they find a security exploit in the hash, or if they brute force the system with a $1/2^{64}$ chance of bypassing the integrity check. \newline

\noindent If there is expected to be simultaneous messages in transit, there could be problems for encryption and decryption. Suppose $CTR_R~=~CTR_S~=~X$ for some $X$, and both devices send each other an encrypted message at the same time. Then, $CTR_R~=~CTR_S~=~X+1$ while the message is midway, and both will attempt to decrypt at $CTR_R~=~CTR_S~=~X+2$. Since the protocol requires a mismatch of $CTR_R$ and $CTR_S$ at decryption time, we could fix this by simply storing two separate counters for encryption and decryption. Instead, suppose both systems have a counter $E\_CTR$ incrementing upon encryption and a counter $D\_CTR$ incrementing upon decryption. Then, $E\_CTR_S = D\_CTR_R$ and $E\_CTR_R = D\_CTR_S$ (except when a message is in transit). Now, since sending and receiving messages increment separate counters, a simultaneous attempt to send a message will not cause the counters to match before decryption is executed.

\section{Superiority of Our Construction}
In this section, we aim to demonstrate either security equivalence or superiority of our simple encryption scheme over other similar hypothetical hash-based encryption protocols.

\subsection{Hashing the Secret Key Instead of a Copy} \label{Hashing SK}
While this may lower the system's SRAM requirement, hashing $SK$ would cause previously used keys to be no longer accessible by any of the communicating parties. Thus, if an encrypted message from an IoT device $S$ were to fail to be delivered to its server $R$, we would have $SK_S~=~H(SK_R)$, and the keys would be desynchronized.

To fix desynchronization in our protocol, the server $R$ could simply test a range of values ahead of its current $CTR_R$. If none of them decrypt, it can conclude the message was illegitimate and reset back to $CTR_R$ to await a legitimate message. However, it cannot reset back to its key value at $CTR_R$ if the original value of $SK$ is lost upon hashing it. So if the protocol hashed $SK$ instead of a copy, there would be no way for $R$ to resynchronize itself in the $SK_S~=~H(SK_R)$ scenario, and no way to communicate to $S$ that it is desynchronized. Thus, $R$ would have to send data until it has hashed more times than $S$. But a server seldom sends more data to an IoT device than vice versa, and therefore $SK_S~=~H(SK_R)$ would be an unrecoverable situation caused by a single packet loss.

Additionally, even when working correctly, $R$ would not be able to verify authenticity of a message, since the original shared secret used to encrypt the message would be destroyed upon being hashed.

\subsection{Using Multiple Hash Functions}
Suppose you implement a scheme that employs multiple hash functions, and decides which one to use before hashing ($SK'\parallel CTR_S$). Then, if any of the hash functions are vulnerable to a passive attack, a fraction of the keys may be deduced by an outsider. Once an outsider can deduce some of the keys, the outsider knows $SK$ and can compute the value of all the keys. Thus, the multiple hash scheme would only be as secure as its least secure hash. For this reason, we elect to use a single, trustworthy hash function to encrypt all the keys.

\subsection{Additional Manipulation of the Input}
Because the hash itself is known to be secure, generation of the secret material used as an input does not need to be any more unpredictable than a counter attached to a shared secret. If there exists an active attack against $H$, then a more unpredictable input in $H$ would not change an outsider's ability to attack the data integrity. Similarly, if there exists a passive attack against $H$, an outsider would still be able to compute the contents of the original message unless the input to $H$ is itself ciphertext. Because using two different encryption protocols to encrypt a single file would be much more heavyweight, we consider this an impractical option for the IoT.

\section{Security Analysis}
Assume that the secret key was placed in both devices securely, and that an outsider cannot find the secret key through any manipulation of the hardware of the IoT device. Also assume that one of the communicating devices is an IoT device, and the other is a server or a Fog node with more available computational resources. We will assume that an outsider can read, manipulate, reroute, and delete messages as they travel to the intended recipient. We also assume that an outsider can send their own, illegitimate messages to either device. For the rest of this paper, we will be using the BLAKE2s hash (from the BLAKE2 hashes) as our choice of hash function.

\subsection{Man-in-the-Middle (MITM) Passive Attack}
An MITM interceptor would only be privy to the encrypted message. Since they do not know the secret key, they cannot compute the decrypted message. Due to the well-known security of the BLAKE2 hashes, they cannot compute $SK$ from knowledge of just the hashes and the current $CTR$ value \cite{BLAKE2}. 

\subsection{Message Modification Attack}
Modification of the encrypted data would cause the BLAKE2s hash to map the data packet to a different and unpredictable 32-byte string upon decryption. Due to the fact that our protocol has 8-bytes (64-bits) for the integrity check, the chance of the integrity check failing is $1/2^{64}$, or once in every $1.8~\times~10^{19}$ attempts. 

\subsection{Replay Attack}
An interceptor which observes the delivery of a legitimate message cannot illegitimately send as many copies of that message to the receiver as they desire. Because the counter increments for each encryption and decryption, each encryption requires a different and unique key to decrypt. And since a change in $CTR$ causes an unpredictable and large change in the hash function’s output, an outsider could not replay messages to the receiver. An outsider can only impersonate a legitimate sender if the outsider knows the secret key, which could then be appended with the correct counter value to be decrypted.

\subsection{Denial-of-Service (DoS) Attack}\label{DOS}
A repeated active message-modification attack on transmitted messages will cause a stream of data to all be correctly interpreted as compromised data. Assuming the server or Fog node is configured to anticipate legitimate messages periodically from the IoT devices, it can detect when a DoS attack is happening with certain devices that have not sent a legitimate message in an unusually large amount of time. Because an outsider cannot impersonate a legitimate message, DoS attacks can be easily detected and addressed on a temporally periodic basis.

\subsection{Counter Desynchronization Attack}
An outsider may attempt to permanently deny service and communication between the devices by deleting a message en route, causing the counters to desynchronize. This could easily be detected and diagnosed, as discussed in Section \ref{DOS}, because it would cause a device to be unable to interpret legitimate messages. As a way to offset the service delay caused by this attack, decryption done by the server can check within a range of the current counter value to attempt to resynchronize.

However, the larger the range the server might check, the more chances an illegitimate message may be decrypted correctly and pass the integrity check. But even with a range of 5,000,000, accounting for the possibility that an outsider has tampered with 265MB of data before the server has had a chance to react, an outsider would still have only a $1/(2.7~\times~10^{13})$ chance of passing an integrity check under any of those 5,000,000 keys. So long as the range chosen is very small compared to $2^{64}$, testing ahead for counter desynchronization should not compromise the security of the system. Note that the server should only check counter values above its current value, so as not to resynchronize to a value that's already been used and create a vulnerability to a replay attack.

While we are aware of methods to ensure synchronization that are computationally simpler (such as sending plaintext $CTR$ values protected by a MAC to the receiver), we are not aware of a simpler method that gives no additional burdens to the IoT device. Since we assume the server is much more computationally powerful than the IoT device, we place the burden of resynchronization solely onto it.

\subsection{Length Extension Attack}
The BLAKE2 hashes are immune to length extension attacks \cite{BLAKE2}. Choosing a hash that is immune to such attacks is necessary for sufficient security.

\section{Memory Requirements}
Using the BLAKE2s hash (107-bytes) and a 32-byte secret key requires 139-bytes of global variable storage. If there is expected to be simultaneous messages in transit, then there would be an additional 5-byte $D\_CTR$ required, increasing the global variable storage requirement to 144-bytes.

$SK'$, a copy of the shared secret and its appropriate counter, is also created in volatile memory to be hashed. This would have the same memory size as the secret key. Hence, the minimum and maximum memory requirements of our scheme are as shown in Equations \ref{eq:6} and \ref{eq:7}:

\begin{equation}
size(H) + 2*size(SK \parallel CTR)
\label{eq:6}
\end{equation}

\begin{equation}
size(H) + 2*size(SK \parallel CTR) + size(CTR)
\label{eq:7}
\end{equation}

In our case, this means 139-bytes (or 144-bytes) are required to store the global variables, and up to 32-bytes at a time to store the local variables, for a total of 171-176 bytes. This accounts for all the SRAM usage except for local variables created to control loops. If a keystream of some length were computed ahead of time, each key would require 32-bytes of memory, and the computation for encryption and decryption would be negligible in terms of both required memory and computation speed, as it would be a simple XOR. The keys could easily be stored in non-volatile memory until use. Because of the reasons given in Section \ref{Hashing SK}, we believe the minimum SRAM requirement for a hash-based encryption protocol should be no less than the calculated $size(H)~+~2*size(SK~\parallel~CTR)$.

\subsection{Measurement Methodology}
We measured the same encryption schemes for both memory and efficiency. For a fair comparison, we chose exclusively 256-bit encryption protocols, to match our 32-byte hashes. We chose to test AES, Speck, and SpeckTiny, a variant of Speck that reduces memory requirements at the cost of performance speed. Notably, among the AES variations we chose to test, we know AES256-EAX to have serious vulnerabilities \cite{EAX}. All XTS schemes tested are used in their Single Key mode, and Speck256Tiny will henceforth be referred to as ``SpeckTiny."

To measure runtime SRAM usage, we printed the SRAM usage at the beginning of the Arduino’s setup function and during each of the necessary operations ($encrypt,~decrypt,~setKey,~and~addAuthData$) as applicable. Reported is the SRAM usage of the most resource-intensive operation of the ones listed above. It is calculated by taking the difference between the SRAM usage before and during an operation.

Idle SRAM usage is the SRAM memory required to simply construct the encryption protocol and initialize it. It includes the necessary variables used for computing the ciphertext. It is calculated by printing the SRAM usage directly after constructing the object. However, 150-bytes are used for the $setup()$ and $loop()$ built-in functions in Arduino, and 153-bytes to import any of the libraries. To account for only the weight of the protocols, we subtract this extra memory usage. The total SRAM requirement is the amount of SRAM that needs to be allocated for the protocol. It is calculated as the sum of the Runtime SRAM and Idle SRAM Measurements. All the code used to assess the memory costs of these protocols is available at \url{https://github.com/matthewchunqed/lightweight-encryption}.

\subsection{Measurement Results}
Table \ref{Table1} shows the SRAM requirements per encryption scheme. Because Arduino devices naturally store all the program's created objects in its embedded SRAM \cite{Arduino}, measuring the memory costs solely on the Arduino's SRAM gives a clear indication of memory requirements for deployment into an IoT device. From Table \ref{Table1}, we see that our hash-based protocol is exceptionally lightweight in terms of its memory requirement. Among the other encryption protocols, we see that all modes of SpeckTiny and AES256-CTR also are particularly lightweight.

\begin{table}[ht]
\centering
\begin{tabular}{ |p{1.7cm}|p{1.7cm}|p{1.7cm}|p{1.7cm}| }
\hline
Encryption Scheme&Runtime Usage& Idle Usage &Total Usage \\
\hline
\cellcolor{lightgray}Proposed & 39 & 137 & 176 \\
\cellcolor{lightgray}AES256-GCM & 304 & 442 & 746 \\
\cellcolor{lightgray}Speck256-GCM   & 394 & 464 & 858 \\
\cellcolor{lightgray}SpeckTiny-GCM  & 160 & 224 & 384 \\
\cellcolor{lightgray}AES256-EAX & 284 & 426 & 710 \\
\cellcolor{lightgray}Speck256-EAX & 378 & 448 & 826  \\
\cellcolor{lightgray}SpeckTiny-EAX & 138 & 208 & 346 \\
\cellcolor{lightgray}AES256-CTR  & 2 & 361 & 363\\
\cellcolor{lightgray}XTS-AES256 & 331 & 339 & 670\\
\cellcolor{lightgray}XTS-Speck256 & 361 & 361 & 722 \\
\cellcolor{lightgray}XTS-SpeckTiny & 121 & 121 & 242\\
\hline
\end{tabular}
\newline
\caption{SRAM Requirements per Encryption Scheme (in bytes)}
\label{Table1}
\end{table}

\section{Performance Tests}
All the code used to test the protocol's efficiency is also available at \url{https://github.com/matthewchunqed/lightweight-encryption}. The Arduino Crypto 0.4.0 library was used to port this protocol onto the devices, and all the experiments comparing our protocol with other protocols was done using its provided example code.

\subsection{Performance Test Methodology}
We coded our proposed scheme in C++ using the Arduino IDE and ran the code on the devices below. We used the Arduino Crypto 0.4.0 library for access to the code. To compare performance speeds with other schemes, we used the example code provided by the Crypto library. Since hash functions normally take inputs of any length and hash them to a fixed size, the code for a hash used in the protocol can be modified to always anticipate inputs of a fixed, 64-byte size which may slightly increase program efficiency. Doing this small modification for BLAKE2s yielded no non-negligible benefit.

\subsection{Performance Test Results}
We tested our protocol on the ESP32-S2-Kaluga-1, the Arduino Zero, and the Arduino Nano. The Kaluga has a high amount of available computational resources, the Nano has a significantly low amount, and the Arduino Zero has a moderate supply. 
\subsubsection{ESP32-S2-Kaluga-1:} 
\begin{figure}[ht]
\centerline{\includegraphics[width=28pc]{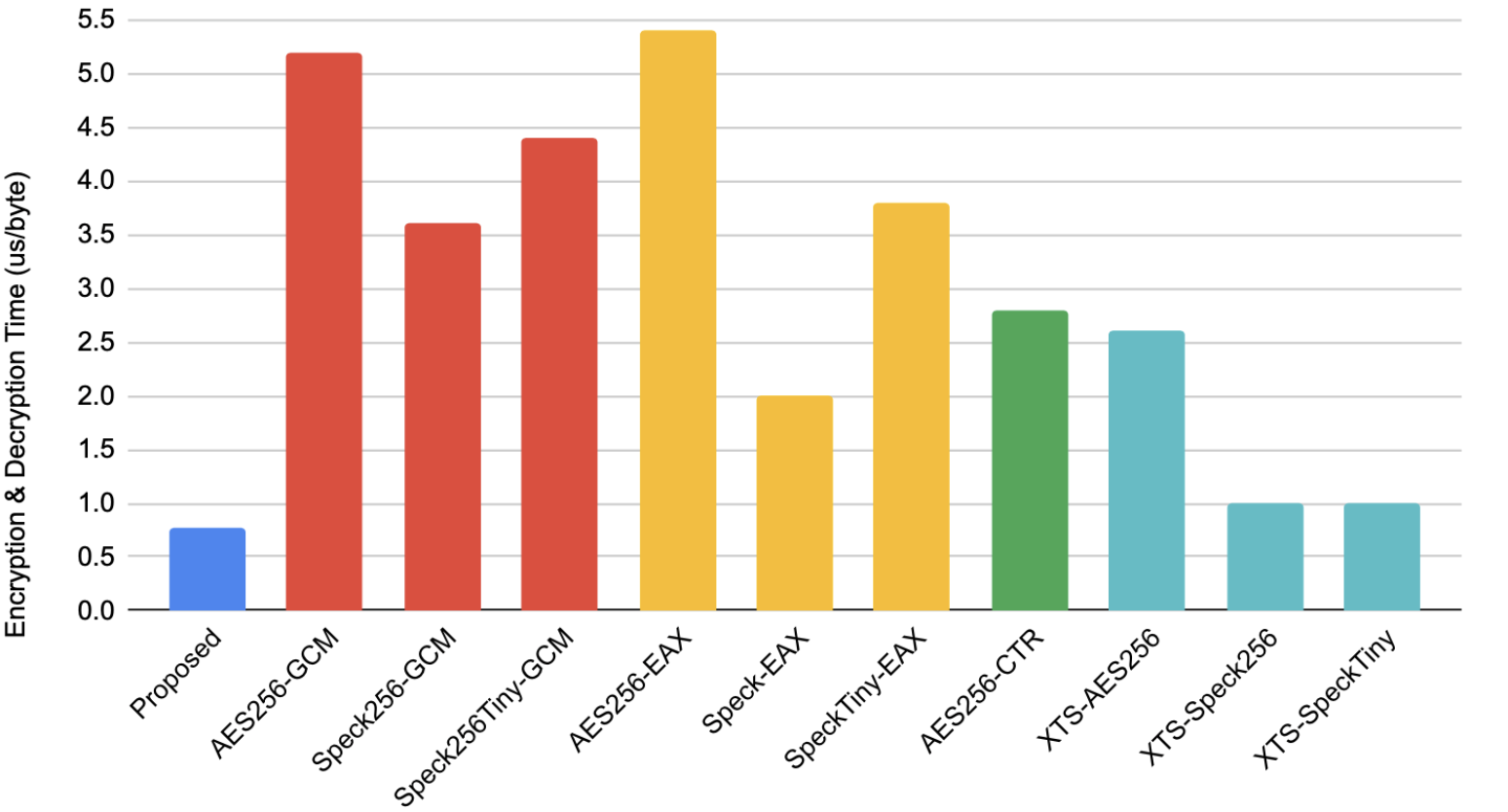}}
\caption{Combined Time to Encrypt and Decrypt per Scheme on ESP32-S2-Kaluga-1}
\label{fig2}
\end{figure}

Table \ref{Table2} shows the performance on a ESP32-S2-Kaluga-1 with a 2.4GHz Clock, 128Kb Flash, and 320Kb SRAM. Figure \ref{fig2} compares the combined encryption and decryption time between encryption schemes on the ESP32-S2-Kaluga-1. In both, we can see that the proposed protocol encrypts faster than any other protocol. Thus, the proposed protocol is the most efficient in terms of both speed and memory on the ESP32-S2-Kaluga-1.

\begin{table}[ht]
\centering
\begin{tabular}{ |p{1.7cm}|p{1.7cm}|p{1.7cm}|p{1.7cm}| }
\hline
Encryption Scheme&Encryption Time ($\mu s/byte$)& Decryption Time ($\mu s/byte$) &Key Setup Time ($\mu s$) \\
\hline
\cellcolor{lightgray}Proposed & 0.39 & 0.39 & N/A \\
\cellcolor{lightgray}AES256-GCM & 2.6 & 2.6 & 41 \\
\cellcolor{lightgray}Speck256-GCM   & 1.8 & 1.8 & 25 \\
\cellcolor{lightgray}SpeckTiny-GCM  & 2.2 & 2.2 & 29 \\
\cellcolor{lightgray}AES256-EAX & 2.7 & 2.7 & 44 \\
\cellcolor{lightgray}Speck256-EAX & 1.0 & 1.0 & 28  \\
\cellcolor{lightgray}SpeckTiny-EAX & 1.9 & 1.9 & 31 \\
\cellcolor{lightgray}AES256-CTR  & 1.4 & 1.4 & N/A\\
\cellcolor{lightgray}XTS-AES256 & 1.3 & 1.3 & 0.7\\
\cellcolor{lightgray}XTS-Speck256 & 0.5 & 0.5 & 11 \\
\cellcolor{lightgray}XTS-SpeckTiny & 0.9 & 0.1 & 1\\
\hline
\end{tabular}
\newline
\caption{Performance on ESP32-S2-Kaluga-1}
\label{Table2}
\end{table}

\subsubsection{Arduino Zero:} 
\begin{figure}[ht]
\centerline{\includegraphics[width=28pc]{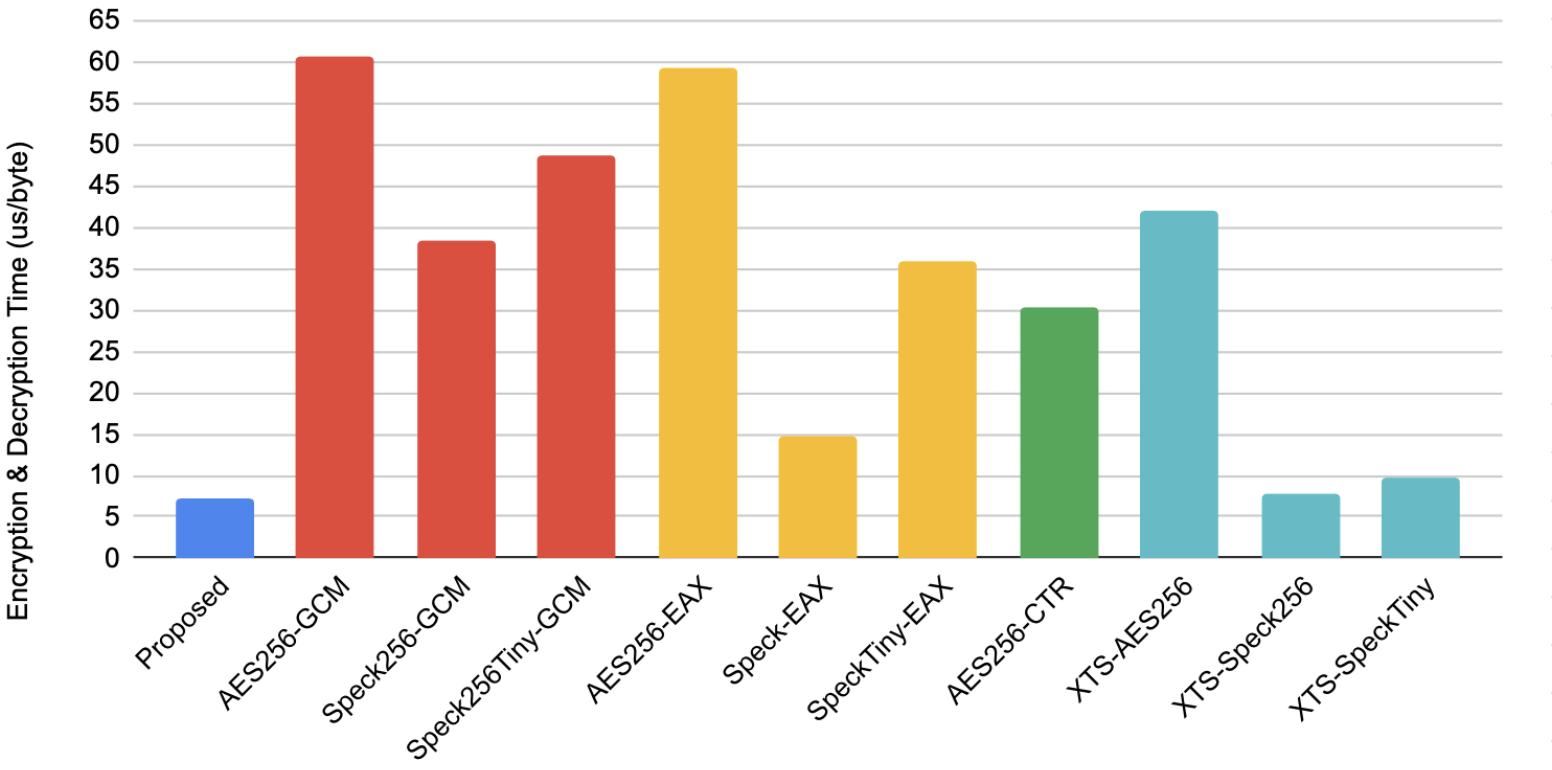}}
\caption{Combined Time to Encrypt and Decrypt per Scheme on Arduino Zero}
\label{fig3}
\end{figure}

Table \ref{Table3} shows the performance on the Ardiuno Zero with a 48MHz Clock, 256Kb Flash, and 32Kb SRAM. Figure \ref{fig3} compares the combined encryption and decryption time between encryption schemes on the Arduino Zero. In both, we can see that the proposed protocol encrypts faster than any other protocol on the Arduino Zero. Thus, the proposed protocol is also the most efficient in terms of both speed and memory on the Arduino Zero.

\begin{table}[ht]
\centering
\begin{tabular}{ |p{1.7cm}|p{1.7cm}|p{1.7cm}|p{1.7cm}| }
\hline
Encryption Scheme&Encryption Time ($\mu s/byte$)& Decryption Time ($\mu s/byte$) &Key Setup Time ($\mu s$) \\
\hline
\cellcolor{lightgray}Proposed & 3.6 & 3.6 & N/A \\
\cellcolor{lightgray}AES256-GCM & 30.3 & 30.3 & 573 \\
\cellcolor{lightgray}Speck256-GCM & 19.2 & 19.2 & 213 \\
\cellcolor{lightgray}SpeckTiny-GCM  & 24.3 & 24.3 & 291 \\
\cellcolor{lightgray}AES256-EAX & 29.6 & 29.6 & 579 \\
\cellcolor{lightgray}Speck256-EAX & 7.4 & 7.4 & 228  \\
\cellcolor{lightgray}SpeckTiny-EAX & 17.9 & 17.9 & 309 \\
\cellcolor{lightgray}AES256-CTR  & 15.2 & 15.2 & N/A\\
\cellcolor{lightgray}XTS-AES256 & 14.8 & 27.3 & 96\\
\cellcolor{lightgray}XTS-Speck256 & 3.7 & 4.2 & 100 \\
\cellcolor{lightgray}XTS-SpeckTiny & 9.0 & 0.8 & 11\\
\hline
\end{tabular}
\newline
\caption{Performance on Arduino Zero}
\label{Table3}
\end{table}

\subsubsection{Arduino Nano:}
\begin{figure}[ht]
\centerline{\includegraphics[width=28pc]{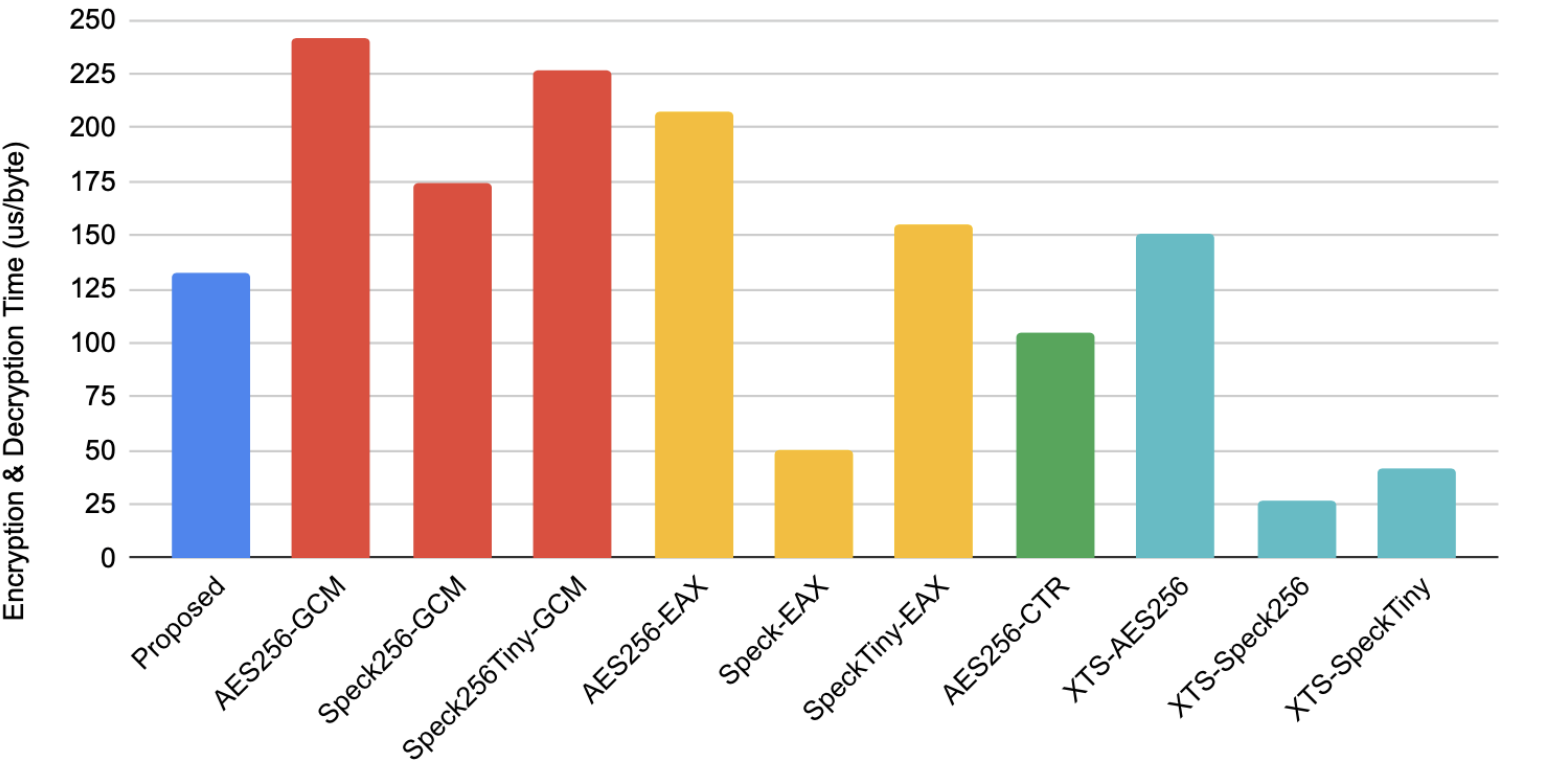}}
\caption{Combined Time to Encrypt and Decrypt per Scheme on Arduino Nano}
\label{fig4}
\end{figure}

Table \ref{Table4} shows the performance on a Arduino Nano with a 16MHz Clock, 32Kb Flash, and 2Kb SRAM. Figure \ref{fig4} compares the combined encryption and decryption time between encryption schemes on the Arduino Nano. In both, we can see that the proposed protocol is not the fastest method of encryption. AES256-CTR, XTS-Speck256, XTS-SpeckTiny, and Speck-EAX encrypt faster than the proposed scheme. However, the proposed scheme is still the most lightweight scheme in terms of memory requirements. Considering both memory requirements and performance efficiency, a developer would likely conclude that the proposed scheme, AES256-CTR, and XTS-SpeckTiny would be the most lightweight and suitable encryption protocols for an appliance built off of the Arduino Nano.

\begin{table}[ht]
\centering
\begin{tabular}{ |p{1.7cm}|p{1.7cm}|p{1.7cm}|p{1.7cm}| }
\hline
Encryption Scheme&Encryption Time ($\mu s/byte$)& Decryption Time ($\mu s/byte$) &Key Setup Time ($\mu s$) \\
\hline
\cellcolor{lightgray}Proposed & 66.5 & 66.5 & N/A \\
\cellcolor{lightgray}AES256-GCM & 126.7 & 126.2 & 1834 \\
\cellcolor{lightgray}Speck256-GCM & 87.3 & 86.8 & 644 \\
\cellcolor{lightgray}SpeckTiny-GCM  & 113.5 & 113.0 & 1223 \\
\cellcolor{lightgray}AES256-EAX & 103.7 & 103.7 & 1886 \\
\cellcolor{lightgray}Speck256-EAX & 25.0 & 25.0 & 695  \\
\cellcolor{lightgray}SpeckTiny-EAX & 77.3 & 77.3 & 1274 \\
\cellcolor{lightgray}AES256-CTR  & 52.4 & 52.4 & N/A\\
\cellcolor{lightgray}XTS-AES256 & 52.4 & 98.7 & 207\\
\cellcolor{lightgray}XTS-Speck256 & 13 & 13.5 & 276 \\
\cellcolor{lightgray}XTS-SpeckTiny & 39.2 & 2.9 & 18\\
\hline
\end{tabular}
\newline
\caption{Performance on Arduino Nano}
\label{Table4}
\end{table}

\subsection{Discussion} 
From the performance tests, we can see the proposed hash encryption protocol is faster than all other encryption schemes, except on the Arduino Nano. Additionally, in terms of memory and computational speed, our lightweight encryption protocol most closely compares to AES256-CTR. 

It is clear from the data tested on the Arduino Nano that even an extremely minimal hash-based encryption scheme is less efficient on some devices than alternatives such as AES256-CTR, or XTS-SpeckTiny. But even being less efficient on the Arduino Nano, our proposed scheme has a smaller memory requirement and is consistently efficient on other devices.

This encryption proposal was made with efficiency and performance in mind. Unlike more heavyweight encryption candidates such as AES-GCM, this protocol is not IND-CCA2 secure, and it gains computational resources for that cost in theoretical security. Still, we believe that the cost to illegitimately decrypt transmitted data from an IoT device using our protocol will far outweigh the value of the data itself. Because of this, we believe this protocol to be a very suitable candidate for general IoT encryption.

Future work could involve the implementation of an HMAC, or other means to increase security of the scheme. Such security upgrades would of course warrant new tests on low-computing power devices, to assess viability in the IoT.

\section{Conclusion}
In this paper we have outlined a scheme that makes use of secure hash functions to implement a flexible, lightweight, and efficient encryption scheme for IoT devices. We believe this framework relying entirely on a secure hash function is suitable for many small devices where computational resources are a premium. While we observed that not all IoT devices are most efficient with such a hash-based encryption scheme, we believe there is a substantial range of small devices which would prefer this encryption setup. Due to its ease to implement and test, we believe it will also be very easy to assess how useful our protocol is for any particular device that may want it. Also, due to its performance and security being entirely dominated by the properties of the hash itself, we believe that innovation in the search for better hash functions will lead to even more lightweight and secure encryption with this same framework.

%
%

\end{document}